\documentclass[%
 aip,
 amsmath,amssymb,
 reprint,%
]{revtex4-1}

\usepackage{graphicx}
\usepackage{dcolumn}
\usepackage{bm}
\usepackage[table,dvipsnames]{xcolor}
\usepackage[utf8]{inputenc}
\usepackage[T1]{fontenc}
\usepackage{natbib}
\usepackage{soul}
\usepackage{ulem}

\usepackage{mathptmx}

\newcommand{\cn}{\text{cn}}
\newcommand{\sn}{\text{sn}}
\newcommand{\dn}{\text{dn}}

\begin{document}

\preprint{AIP/123-QED}

\title[Nonlinear waves in flexible mechanical metamaterials]{Nonlinear waves in flexible mechanical metamaterials}

\author{B. Deng}
 \email{boleideng@g.harvard.edu}
 \affiliation{Harvard John A. Paulson School of Engineering and Applied Sciences, Harvard University, Cambridge, MA 02138}%
\author{J. R. Raney}
\email{raney@seas.upenn.edu}
\affiliation{Department of Mechanical Engineering and Applied Mechanics, University of Pennsylvania, Philadelphia, PA 19104}
\author{K. Bertoldi}%
 \email{bertoldi@seas.harvard.edu}
\affiliation{Harvard John A. Paulson School of Engineering and Applied Sciences, Harvard University, Cambridge, MA 02138}%

\author{V. Tournat}
\email{vtournat@univ-lemans.fr}
\affiliation{Laboratoire d'Acoustique de l'Universit\'e du Mans (LAUM), UMR 6613, Institut d'Acoustique - Graduate School (IA-GS), CNRS, Le Mans Universit\'e, France}%

\date{\today}

\begin{abstract}
Flexible mechanical metamaterials are  compliant structures engineered to achieve unique properties via the large deformation of their components. While their static character has been studied extensively, the study of their dynamic properties is still at an early stage, especially in the nonlinear regime induced by their high deformability. Nevertheless, recent studies show that these systems provide new opportunities for the control of large amplitude elastic waves. Here, we summarize the 
recent results on the propagation of nonlinear waves in flexible elastic metamaterials, and 
highlight possible new research directions.
\end{abstract}

\maketitle

\section{Introduction}

Over the last two decades, metamaterials -- materials whose properties are defined by their structure rather than their composition -- have been a real magnet for scientists, generating significant interest in the research community~\cite{engheta2006metamaterials,cui2010metamaterials,craster2012acoustic,deymier2013acoustic}. While initial efforts focused on metamaterials that manipulate electro-magnetic~\cite{engheta2006metamaterials}, acoustic~\cite{craster2012acoustic,deymier2013acoustic} or thermal~\cite{maldovan2013} properties, in recent years the concept has also been extended to 
mechanical systems~\cite{BertoldiK.2017Fmm,KadicMuamer2013Mbe,ChristensenJohan2015Vtfm}. Ongoing advances in digital manufacturing technologies
~\cite{spadaccini2019,clausen2015,raney2015,sundaram2019} have stimulated the design of mechanical metamaterials with highly unusual properties, including negative Poisson's ratio~\cite{LAKES_1987,Bertoldi2010AdvMat}, negative thermal expansion~\cite{Wang_2016,Boatti_2017}, and negative compressibility~\cite{Baughman_1998} in the static regime as well as low-frequency spectral gaps~\cite{Liu2000,Foehr_2018}, negative dynamic properties~\cite{Ding_2007}, and advanced dispersion effects~\cite{Ma_2018} in the dynamic regime. Further, it has been shown that large deformations and mechanical instabilities can be exploited to realize flexible mechanical metamaterials (flexMM) with new modes of functionality~\cite{BertoldiK.2017Fmm}. The complex and programmable deformation of flexMMs make them an ideal platform to design reconfigurable structures~\cite{haghpanah2016multistable}  as well as soft robots~\cite{Rafsanjani_2019} and mechanical logic devices~\cite{Raney2016,bilal2017,jiang2019bifurcation}. 
Further, they also provide opportunities to manipulate the propagation of finite amplitude elastic waves. Differently from granular media whose nonlinear
response is determined by the contacts between grains~\cite{nesterenko2013dynamics,Tournat_2010,theocharis2013nonlinear}, the nonlinear dynamic response of flexMMs is governed by their architecture. By carefully choosing the geometry, a flexMM can be designed to be either monostable or multistable or to support large internal rotations -- all features that have been shown to result in interesting non-linear dynamic phenomena.  

In  this perspective  article, we  first review the nonlinear dynamic effects that have been recently reported for flexMMs: the propagation and manipulation of vector elastic solitons, rarefaction solitons, and topological solitons (also referred to as transition waves). We then describe the numerical and analytical tools that are typically used to investigate the propagation of these nonlinear waves. Finally, we outline the key challenges and opportunities for future work in this exciting area of research.

\section{Nonlinear dynamic effects in flexMM}
\label{sec:effects}

While nonlinear elastic waves in engineered materials have mostly been experimentally studied in granular media~\cite{nesterenko2013dynamics,Tournat_2010,boechler2011,theocharis2013nonlinear,Chong_2017,allein2020linear}, flexMM also provide an ideal environment for their propagation, since they can support a wide range of effective nonlinear behaviors that are determined by their architecture. By carefully tuning these nonlinear behaviors, novel dynamic effects have been demonstrated. First, metamaterials based on the rotating square mechanism have been shown to support the propagation of elastic vector solitons - solitary pulses with both translational and rotational components, which are coupled together and copropagate without distortion nor splitting due to the perfect balance between dispersion and nonlinearity \cite{Deng_2017,Deng_2018NM,deng2019focusing}.
Second, since flexMM can typically support tensile deformation, the propagation of rarefaction solitons has also been demonstrated \cite{deng2019propagation,yasuda2019origami,Fraternali_2014,Deng_2020}. Third, by designing their energy landscape to be multiwelled, it has been shown that they can support the propagation of topological solitons (also referred to as transition waves) - nonlinear pulses that sequentially switch the structural elements from one stable state to another \cite{nadkarni2016unidirectional,Raney2016,vasios2021universally,zareei2020harnessing,jin_2020}. 

\begin{figure*}
\includegraphics[width=\textwidth]{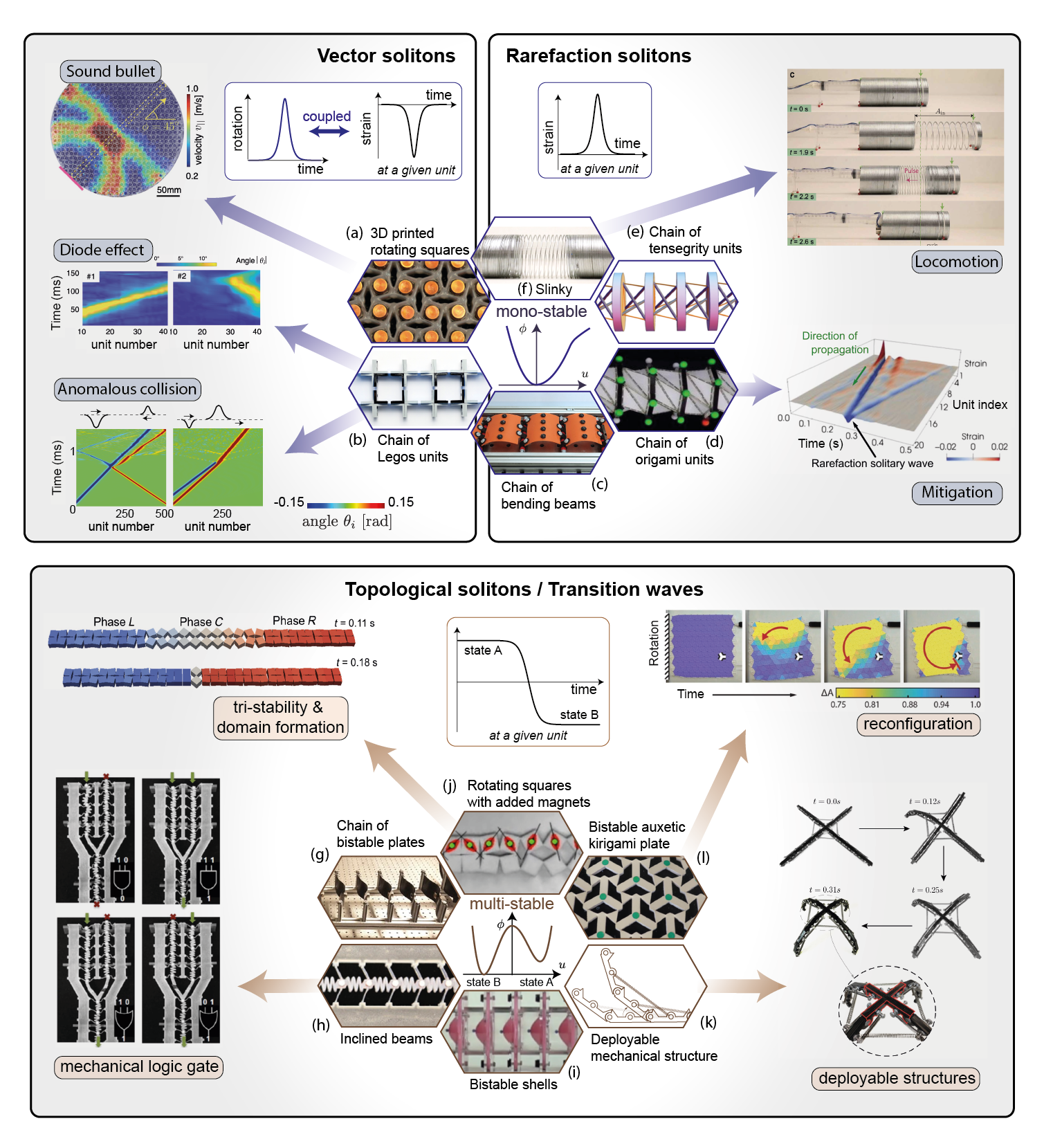}
\caption{\label{fig1} FlexMMs    provide   a  rich platform  to  manipulate  the  propagation  of  nonlinear  waves. Vectors solitons have been observed in (\textbf{a}) a 2D flexMM based on the rotating squares mechanism~\cite{Deng_2017} (\textbf{b}) a chain of Lego units connected by flexible hinges~\cite{Deng_2018NM}.  Rarefaction  solitons have been observed in (\textbf{c}) a  chain of hinged buckled beams~\cite{deng2019propagation}, (\textbf{d}) a  chain of origami units (graph and picture licensed under a Creative Commons Attribution (CC BY-NC) license, reproduced and cropped from~\cite{yasuda2019origami}), (\textbf{e}) a  chain of tensegrity units (reproduced from~\cite{Fraternali_2014} with the permission of AIP Publishing)  and (\textbf{f})~a Slinky~\cite{Deng_2020}. Topological solitons (transition waves) have been observed in (\textbf{g}) a chain of bistable plates coupled by magnetic force~\cite{nadkarni2016unidirectional,Hwang_2018SR} (picture licensed under a Creative Commons Attribution (CC BY) license, reprinted and cropped from Ref.~\cite{Hwang_2018SR}), (\textbf{h}) a  chain of bistable inclined beams coupled by elastic elements~\cite{Raney2016}, (\textbf{i}) a  chain of bistable shells coupled by pressurized air~\cite{vasios2021universally}, (\textbf{j})~a system of rotating squares with embedded magnets~\cite{Yasuda_2020}, (\textbf{k}) a 1D linkage~\cite{zareei2020harnessing},  (\textbf{l}) a 2D multistable kirigami structure~\cite{jin_2020}.
}
\end{figure*}

\subsection{Elastic vector solitons}

Flexible metamaterials comprising a network of squares connected by thin and highly deformable ligaments (see Fig.~\ref{fig1}(a) and (b)) have long attracted significant interest due to their effective negative Poisson's ratio~\cite{Grima2000,Cho2014,SShan2015} and their support of buckling-induced pattern transformations~\cite{Mullin2007,Bertoldi2010AdvMat}. 
Additionally, it has recently been shown via a combination of experiments and analyses that the nonlinear dynamic response of these structures is also very rich~\cite{Deng_2017,Deng_2018NM,deng2019anomalous,deng2019focusing}. First, it has been demonstrated that even one-dimensional (1D) chains of these metamaterials support the propagation of elastic vector solitons with both translational and rotational components which are coupled together and copropagate without dispersion~\cite{Deng_2017,Deng_2018NM}. The existence of these pulses is enabled by the perfectly balanced dispersive and nonlinear effects~\cite{dauxois2006}.  While vector solitons have been previously reported in optics~\cite{Tang_2008}, their observation in networks of hinged squares is the first  for the elastic case. Importantly,  the vectorial nature of such solitons gives rise to a vast array of exotic mechanical phenomena. For example, due to the weak coupling between their two components, at small enough amplitudes the vector solitons become dispersive and fail to propagate~\cite{Deng_2018NM}. Further, the vectorial nature of the supported solitons leads to anomalous collisions~\cite{deng2019anomalous}. While, as expected, the solitons emerge unaltered from the collision if they excite rotations of the same direction, they do not penetrate each other and instead repel one another if they induce rotations of the opposite direction. Finally, it has been shown that nonlinear propagation in two-dimensional (2D) systems of rotating squares exhibit very rich direction-dependent behaviors such as the formation of \textit{sound bullets} and the separation of pulses into different solitary modes~\cite{deng2019focusing}. 
As such, these studies suggest that flexible metamaterials based on the rotating-square mechanism may represent a powerful platform to manipulate the propagation of nonlinear pulses in unprecedented ways.

\subsection{Rarefaction solitons}

Granular systems that derive their nonlinearity from Hertzian contact are well known to support the propagation of compressive solitons in many instances~\cite{sen2008solitary,Shen2014,spadoni2010,Nesterenko2005,Hong2002,daraio2006tunability,nesterenko1983}. Yet it is challenging for such systems to support rarefaction solitons due to their lack of tensile cohesion. While it has been shown via a combination of theoretical and numerical analyses that a precompressed discrete chain with strain-softening interactions could support rarefaction solitons~\cite{herbold2013propagation}, experimental demonstration of this behavior has remained elusive due to the challenges in fabricating an effective strain-softening mechanism. By contrast, flexMMs can be easily designed to support softening nonlinearity under compression and, therefore, rarefaction solitons, an effect that can in principle allow useful applications by enabling efficient impact mitigation. This is the case of a 1D array of buckled beams~\cite{deng2019propagation} (Fig.~\ref{fig1}(c)), as well as a chain of triangulated cylindrical origami~\cite{yasuda2019origami} (Fig.~\ref{fig1}(d)). Both systems exhibit effective strain-softening behavior and have been shown to support the propagation of rarefaction solitons. Further, rarefaction solitons have been predicted but not yet experimentally observed in tensegrity structures~\cite{Fraternali_2014} (Fig.~\ref{fig1}(e))  and statically compressed metamaterials based on the  rotating-square mechanism~\cite{deng2018effect}. Note that, for the latter, vector rarefaction solitons are predicted to propagate, sharing the interesting features discussed in the previous section.

Beyond impact mitigation, rarefaction solitons have also been harnessed to enable locomotion in a slinky-based soft robot~\cite{Deng_2020} (see Fig.~\ref{fig1}(f)). The nondispersive nature and compactness of the solitary pulses make them
extremely efficient in transferring the energy provided by the actuator to motion, ultimately resulting in an efficient pulse-driven locomotion.

\subsection{Topological solitons / Transition waves}
\label{sec:topsol}

In addition to vector and rarefaction solitons another category of nonlinear wave, comprising what are called transition waves or topological solitons, has also received a significant amount of recent attention. These waves represent moving interfaces that separate regions of different phases and play a major role in a wide range of physical phenomena, including  damage propagation in solids~\cite{cherkaev2005transition},  dynamic phase transitions~\cite{truskinovsky2004origin,Truskinovsky_2006,Truskinovsky_2008} and phase transformations in crystalline materials~\cite{Bhattacharya2003,porter2009phase,falk1984ginzburg,fiebig2016evolution}.
Recently, it has been shown that transition waves can also propagate in flexMM made with elements possessing multi-well energy landscapes, with each energy well corresponding to a stable spatial configuration. When a transition wave propagates in these systems it can be visualized as a solitary pulse that sequentially switches the individual units of the metamaterial from {one stable configuration to another}~\cite{nadkarni2016unidirectional,zheng2019piezo,fang2017asymmetric,jin_2020,shan2015multistable,dorin2019vibration,kidambi2017energy,Raney2016,zareei2020harnessing}. 

Transition waves were first experimentally observed in flexMM in a system comprising  a 1D array  of bistable and pre-stressed composite shells coupled by magnetic force~\cite{nadkarni2016unidirectional} (Fig.~\ref{fig1}(g)). More specifically, the shells are designed to have two energy minima of different height. Therefore, the transition between the two stable states involves a net change in stored potential energy, 
which, depending on the direction of the transition, either absorbs energy or releases stored potential energy. If the bistable shells are initially set to their higher-energy  stable configuration, a sufficiently large displacement applied to any of them can cause the element to transition states, producing a nonlinear transition wave that propagates indefinitely outward from the point of initiation with constant speed and shape.  Further,  transition waves  are not sensitive to the specific signal that triggers them and can be initiated by  any large enough input signal. Such robustness has  been recently harnessed to  concentrate, transmit and harvest  energy independently from the
excitation~\cite{Hwang_2018SR} (see Fig.~\ref{fig1}(g)). Specifically, the energy carried by the transition waves has been {focused and subsequently harvested in lattices by introducing engineered defects and integrating electromechanical transduction}~\cite{Hwang_2018SR}.

Interestingly, because of the energy released upon transition between states  by each element, stable and long-distance propagation of transition waves in multistable systems is possible even in the presence of significant dissipation\cite{Raney2016} - a feature that has been demonstrated for a soft structure composed of elastomeric bistable beam elements connected by elastomeric linear springs. Such ability to transmit a mechanical signal over long distances with high fidelity and controllability has been shown to provide opportunities for  signal processing, as demonstrated by the design of soft mechanical diodes and logic gates~\cite{Raney2016} (see Fig.~\ref{fig1}(h)). Notably, these systems have been also recently realized at the micro-scale using two-photon stereolithography~\cite{song2019additively}, providing a first step towards mechanical chips. However, it is important to note that, while  bistable unit cells with two stable states of different energy levels enable long-distance propagation of transition waves, they inherently prevent bidirectional signal transmission~\cite{nadkarni2016unidirectional}. Further, they require an external source of energy to be provided to reset them to their higher-energy state if additional propagation events are desired. 

Bidirectional propagation of transition waves can be achieved by utilizing bistable elements that possess  equal energy minima~\cite{vasios2021universally} (Fig.~\ref{fig1}(i)). However, since such bistable elements do not release energy when transitioning between their two stable states, the distance traveled by the supported transition waves is limited by unavoidable dissipative phenomena. To overcome this limitation two strategies have been proposed. On the one hand, it has been shown that the propagation distance of transition waves can be extended by introducing elements with tunable energy landscape, since they can be easily set to release the energy required to compensate for dissipation~\cite{vasios2021universally}. On the other hand, long-distance propagation of transition waves has been demonstrated in a 1D array of bistable elements with monotonically decreasing energy barriers~\cite{Hwang_2018}, but such gradient in energy landscape prevents bidirectionality.

FlexMM can also be designed to possess more than two energy minima (Fig.~\ref{fig1}(j)). Just as in the bistable systems described above, transition waves can propagate when a transition from one stable well to another is initiated. However, since multiple types of energetically-favorable transitions are possible (e.g., a system in a higher energy well might support transition waves to two different lower energy wells, each associated with distinct spatial configurations), incompatible transition waves can propagate and collide, leading to non-homogeneous spatial configurations. For example, transition waves have been demonstrated in rotating-square systems with permanent magnets added to the faces~\cite{Yasuda_2020}. In contrast with the buckled elements described above, each unit in the metamaterial supports up to three stable configurations, enabled by the ability of the squares to be stable in `open', `clockwise', or `counterclockwise' configurations.  The ability of multistable systems to support the formation of many configurations of stationary domain walls could allow the design of transformable mechanical metamaterials that can be reversibly tuned across a large range of mechanical properties.

Finally, while all initial studies on the propagation of transition waves in flexMM have considered 1D  chains, recently transition waves have also been studied in flexMMs with higher dimensions. As a first step in this direction, the response of a network of 1D mechanical linkages that supports the propagation of transition waves has been investigated~\cite{zareei2020harnessing} (Fig.~\ref{fig1}(k)). It has been shown that, if the connections between the linkages are properly designed to preserve the integrity of the structure as well as to
enable transmission of the signal through the different components, transition
waves propagate through the entire structure and transform the initial architecture. Further, the propagation of transition waves has also been demonstrated in 2D multistable  elastic kirigami sheets~\cite{jin_2020} (Fig.~\ref{fig1}(l)). While homogeneous architectures result in constant-speed
transition fronts,  topological  defects can be introduced to manipulate the pulses and  redirect or pin transition waves, as well as to split, delay, or merge propagating wave fronts.  

The results discussed in this Section point to the rich dynamic responses of flexMMs. However, in order to enable such interesting behaviors the geometry of flexMM has to be carefully chosen. As such, it is crucial for the advancement of the field to  develop   models that can accurately predict these nonlinear behaviors and their dependency from geometric parameters and loading conditions.

\section{Modeling the nonlinear dynamic response of flexMM}

\begin{figure}
\centering
\includegraphics[width=0.8\columnwidth]{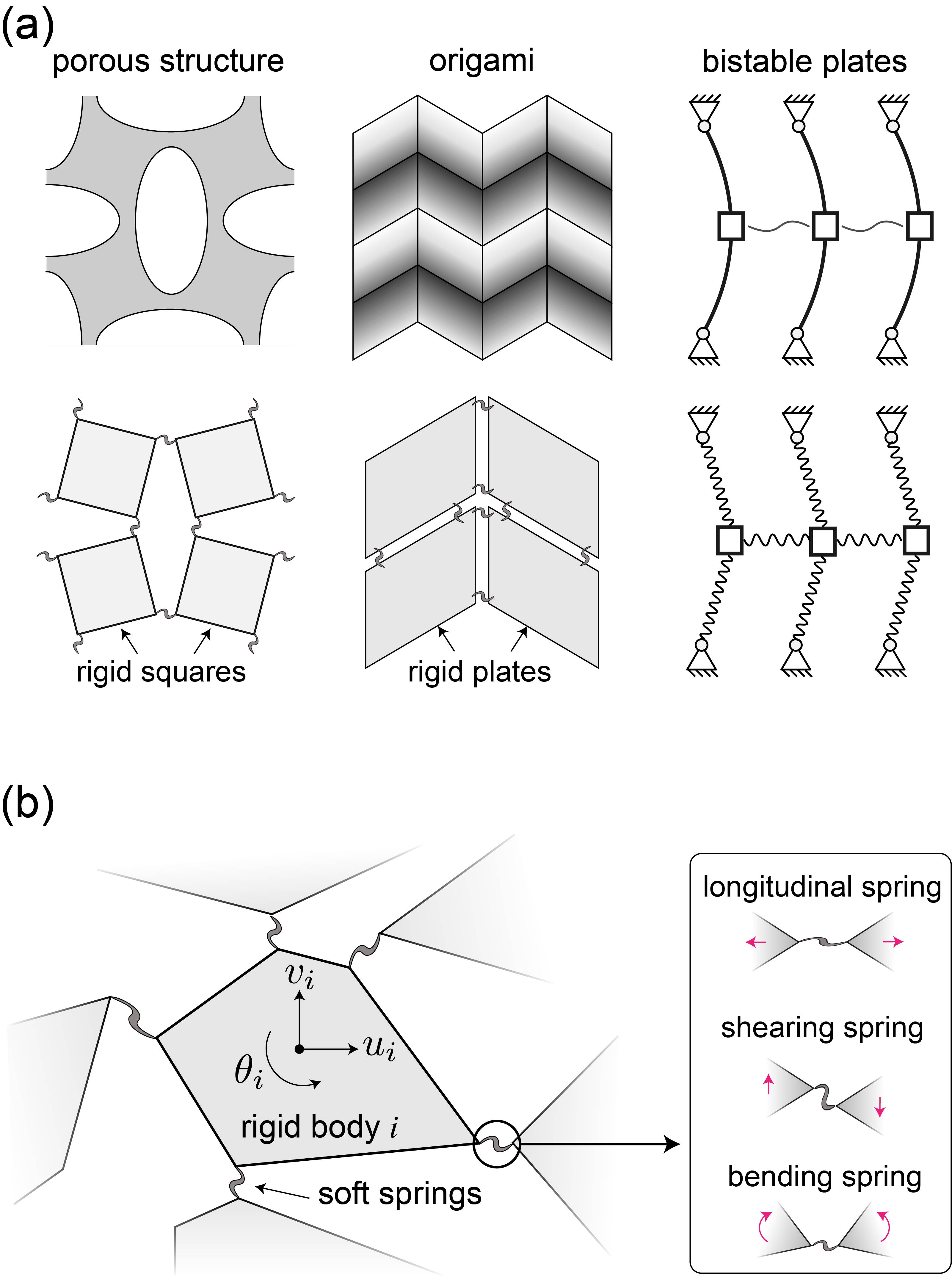}
\caption{\label{fig:modeling}  Modelling the nonlinear dynamic response of FlexMM. (a)~Schematics of three classes of flexMM and their corresponding discrete model. (b) The discrete models typically comprise  networks of rigid bodies connected at  by a combination of rotational and longitudinal  linear springs. }
\end{figure}

Discrete models have traditionally played an important role in  unraveling the dynamic response of structures. Networks of point masses connected by linear springs have been routinely used to understand the propagation of linear waves in solid media~\cite{hussein2014}. Further, by introducing nonlinear springs these models have also enabled investigation of nonlinear waves in engineered media, including granular systems~\cite{nesterenko2013dynamics,Tournat_2010,boechler2011,theocharis2013nonlinear,Chong_2017,allein2020linear} and mass-spring lattices~\cite{dauxois2006}. 
In recent years discrete models have also proven useful to describe the nonlinear dynamic response of flexMM~\cite{Deng_2017,Deng_2018NM,deng2019focusing,deng2018effect,deng2019propagation,Nadkarni_2014,dauxois2006,zareei2020harnessing,Deng_2020,deng2019anomalous,Yasuda_2016,Fraternali_2014,frazier2017atomimetic,Kochmann_2017,Ramakrishnan2020}, as they typically comprise stiffer elements connected by flexible hinges. The stiffer elements are modeled as rigid plates, whereas the response of the hinges is captured using a combination of rotational and longitudinal springs (see Fig.~\ref{fig:modeling}(a)). Note that, since the rotation of the stiff elements plays a crucial role in flexMM, the rotational degrees of freedom of the rigid bodies play an important role in these models. For a typical 2D flexMM, three degrees of freedom (DOFs) are assigned to the $i$-th rigid element: the displacement in $x$ direction,  $u_i$, the displacement in  $y$ direction, $v_i$, and rotation around the $z$ axis, $\theta_i$ (see Fig.~\ref{fig:modeling}(b)). Using these
definitions, the equations of motion for the $i$-th rigid element are given by 
\begin{equation}
\begin{split}
\label{general1}
m_i\ddot{u}_i= \sum_{p=1}^{N_{vi}}F^{u}_{i,p},\;\;\;m_i\ddot{v}_i= \sum_{p=1}^{N_{vi}}F^{v}_{i,p},\;\;\;\text{and}\;\;\;J_i\ddot{\theta}_i = \sum_{p=1}^{N_{vi}}M^\theta_{i,p},
\end{split}
\end{equation}
where $m_i$ and $J_i$ are its mass and moment of inertia, respectively, and $N_{vi}$ denotes its number of vertices. Moreover, $F^{u}_{i,p}$ and $F^{v}_{i,p}$ are  the forces along the $x$ and $y$ directions generated at the $p$-th vertex of
the $i$-th units unit by the springs and $M^{\theta}_{i,p}$ represent the corresponding moment. Note that these forces can be expressed as a function of the DOFs of neighboring elements and are typically calculated assuming linear springs. Unlike typical mass-spring models previously used to investigate nonlinear waves, linear springs are sufficient to capture the dynamic response of flexMM, since the nonlinear behavior comes mainly from geometry. 
Eq.~(\ref{general1}) can subsequently be numerically integrated to obtain the dynamic response of the system. Importantly, these models provide a direct relation to the geometry of flexMMs, thus providing essential insights on their dynamic response.  

For the special case of planar waves with characteristic wavelengths much larger than the unit cells, analytical solutions can also be obtained by taking the continuum limit of the discrete equations of motion~\cite{Deng_2017,Deng_2018NM,deng2019focusing,deng2018effect,deng2019propagation,yasuda2019origami,Nadkarni_2014,dauxois2006}. 
There have been several examples of this:  the response of an array of magnetically coupled bistable plates can be captured by a nonlinear Schr\"{o}dinger equation~\cite{Nadkarni_2014,dauxois2006}; the response of a flexMM based on the rotating-square mechanism can be captured by the nonlinear Klein-Gordon equation \cite{Deng_2017,Deng_2018NM,deng2019focusing,deng2018effect}; the response of a chain of buckled beams follows the Boussinesq equation \cite{deng2019propagation}; and the dynamics of origami chains have been found to follow a Korteweg - de Vries equation \cite{yasuda2019origami}. Depending on the specific geometries of the flexMM and the driving input, these equations, when fully integrable, yield analytical solutions describing solitons \cite{Deng_2017,Deng_2018NM,deng2019focusing,Nadkarni_2014}, rarefaction solitons \cite{yasuda2019origami,deng2019propagation}, and topological solitons \cite{Nadkarni_2014,deng2018effect}. Interestingly, these solutions provide a direct relation of these phenomena to the geometrical parameters of flexMMs. Therefore, they not only allow interpretation of the experimentally and numerically observed phenomena, but also provide opportunities for the rational design of flexMM with targeted nonlinear dynamic responses.


For example, for a metamaterial based on the rotating square mechanism by taking the continuum limit of Eq. (\ref{general1}), retaining
nonlinear terms up to the third order and introducing the traveling wave coordinate  $\zeta = x \cos\phi + y\sin\phi - ct$ (where $x$ and $y$ are the Cartesian coordinates, $t$ indicates time and $\phi$ and $c$ represent direction and velocity of the propagating planar wave) in the governing equations of motion, it is found that the propagation of large amplitude planar waves is described by a nonlinear Klein-Gordon equation of the form~\cite{Deng_2017,Deng_2018NM,deng2019focusing,deng2018effect},
\begin{equation}
    \label{KG}
    \frac{\partial^2 \theta}{\partial\zeta^2}= C_1 \theta + C_2 \theta^2 + C_3 \theta^3 + \mathcal{O}(\theta^4),
\end{equation}
where $C_1$, $C_2$, and $C_3$  are parameters that depend  on the geometry of the flexMM and the flexibility of its hinges   and can therefore be tailored by tuning the metamaterial design. 
Eq.~(\ref{KG}) admits well-known solitary wave solutions
of the form~\cite{Deng_2018NM}:
\begin{equation}
    \label{sol}
    \theta(\zeta) = \frac{1}{D_1\pm D_2\cosh(\zeta/W)} ,
\end{equation}
with
\begin{equation}
    D_1 = -\frac{C_2}{3C_1},\;\;\;D_2 = \sqrt{\frac{C_2^2}{9C_1^2}-\frac{C_3}{2C_1}},\;\;\;\text{and}\;\;\;W = \frac{1}{\sqrt{C_1}} .
\end{equation}
Eq. (\ref{sol}), depending on the sign of $D_1$ and $D_2$, captures different types of stable  non-linear pulses, including solitons, rarefaction solitons \cite{Deng_2018NM} and topological solitons \cite{deng2018effect}.
Moreover, it is important to note that under the assumption of traveling wave coordinates, other types of governing non-linear equations, including the Boussinesq equation \cite{deng2019propagation} and Korteweg-de Vries equation \cite{deng2019anomalous}, can be transformed into a nonlinear Klein-Gordon equations,  making Eq. (\ref{KG}) quite general.

Finally, energy balance considerations have also proven useful to predict the characteristics of topological solitons propagating in dissipative media~\cite{nadkarni2016universal,Hwang_2018}. Specifically, it has been shown that the wave speed can be estimated by balancing the total transported kinetic energy, the difference  between the higher and lower energy wells for the asymmetric elements, and the energy dissipated.   Further, energy considerations can also provide insight into  topological solitons-based energy harvesting~\cite{Hwang_2018SR}.

\begin{figure}
\includegraphics[width=0.9\columnwidth]{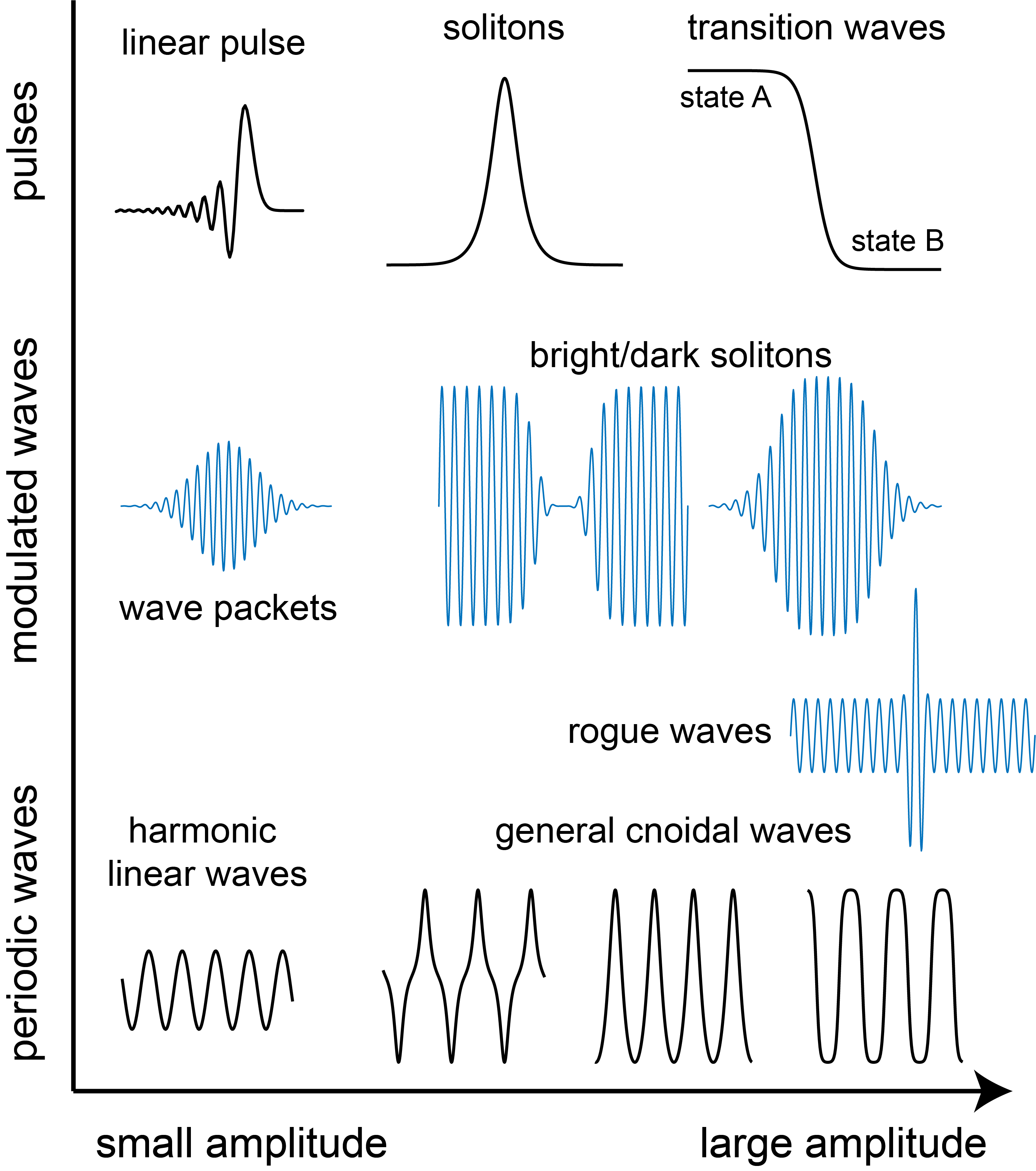}
\caption{A diagram showing different types of linear and nonlinear waves. \label{fig:cnoidal}}
\end{figure}

\section{Outlook}

In summary, this perspective paper has attempted to demonstrate that flexible mechanical metamaterials provide a rich platform to manipulate the propagation of nonlinear waves. We close this paper by
identifying several challenges for future work.

\textbf{Towards nonlinear periodic waves.} Beyond pulse-like, large-amplitude waves with finite spatial and temporal extent  (e.g., solitons and transition waves discussed in this work), it has been shown that rotating-square systems with either quadratic~\cite{mo2019cnoidal} or cubic nonlinearity~\cite{deng2021dynamics} also support the propagation of \textit{cnoidal waves} (see Fig.~\ref{fig:cnoidal}). Cnoidal waves are described by the Jacobi elliptic functions $\dn(\cdot|k)$, $\sn(\cdot|k)$, and $\cn(\cdot|k)$ where $k$ is the elliptic modulus controlling the shape of the elliptical functions. These cnoidal wave solutions extend from linear waves (for
$k\rightarrow 0$) to solitons (for
$k\rightarrow 1$), while covering also a wide family of nonlinear periodic waves~\cite{deng2021dynamics}. 
Furthermore, as depicted in Fig.~\ref{fig:cnoidal}, flexMM could also provide a laboratory test bed for the observation of other types of nonlinear waves~\cite{Yang_2020,Yasuda_2020_data}, including bright/dark solitons~\cite{Nadkarni_2014}, breathers, and rogue waves~\cite{Onorato_2013} (large-amplitude waves that suddenly appear/disappear unpredictably, typically observed in surface water waves~\cite{Chabchoub_2011}). Harmonic generation, frequency conversion\cite{Guo2018,Guo2019} and even actuation (via effects such as frequency-down conversion) are other exciting possibilities to be explored by the rational design of the nonlinear properties of flexMM. Typically in experiments, the periodic and modulated waves depicted in Fig.~3 are generated by a low-frequency shaker driving a boundary of the FlexMM in the range 10~Hz - 10 kHz. Other types of transducers, drivers or actuators are conceivable depending on the dimensions of the microstructure, the frequency range of interest and the desired effects.

\textbf{New flexMM designs.} So far, the nonlinear dynamic response of a limited number of flexMM designs has been investigated. FlexMM based on origami, kirigami, tensegrity structures, and rotating-units other than squares (e.g. triangles or hexagons) may provide additional opportunities to manipulate the propagation of nonlinear waves.  Also, the nonlinear dynamic responses of three-dimensional architectures remain largely unexplored, and may open new avenues for wave management.  

\textbf{Going beyond periodic systems.} While most previous studies have focused on the propagation of nonlinear pulses in periodic and homogeneous structures, new opportunities may arise when investigating the interactions of large-amplitude waves with free surfaces, inhomogeneous structures, and sharp interfaces. How do the nonlinear pulses propagate along free surfaces? What is the effect of internal interfaces on soliton propagation? How do other spatial variations such as gradients in initial angle, gradients in mass, or gradients in the stiffness of the hinges affect the propagation of waves through the material? Can these be used to steer a beam or otherwise affect an incident plane wave? All these questions remain unanswered.

\textbf{Targeted nonlinear dynamical responses.} While the focus so far has been on the development of tools to predict and characterize the propagating nonlinear waves in flexMM, an important question that is still unanswered is:  How should one design the structure, including unit cell geometry, inhomogeneities such as gradients and interfaces, etc., to enable a target dynamic response?  Target dynamic responses may include highly efficient damping for impact mitigation; optimal wave guiding (i.e., optimal energy confinement and propagation along a determined path); and lensing of solitons for optimal energy concentration. To allow the automated design of flexMM architectures that are optimal for achieving a specified set of target dynamic properties, one could couple discrete models with machine learning algorithms, such as neural networks and deep learning.

\textbf{Control of nonlinear waves on the fly.} 
Since the characteristics (i.e., shape, velocity, and amplitude) of nonlinear waves propagating through FlexMM can be tuned by varying the nonlinear response of the underlying medium, which can be effectively altered by (locally) deforming the metamaterial, we envision that the applied deformation could be a powerful tool to manipulate the pulses. Local deformations applied to the FlexMM could provide a {mechanism} to change the characteristics as well as the path  of the propagating pulses on the fly. This could provide opportunities for time-space modulation of the propagating pulses~\cite{Ramakrishnan2020},  real-time  control of  waves, and tunable non-reciprocal transmission~\cite{Nassar_2020}. Further, since collisions of solitons in flexMM may result in anomalous interactions that provide opportunities to remotely detect, change, or {eliminate} high-amplitude signals and
impacts~\cite{deng2019anomalous},  we envision the use of collisions to pave new ways toward the advanced control of large amplitude mechanical
pulses. 

\textbf{Reconfigurability via transition waves.}  As described in Section~\ref{sec:topsol}, the propagation of  topological solitons (transition waves) in multi-stable flexMM can reconfigure all or part of the sample. Since this reconfiguration can be initiated even by a localized, weak impulse, 
a number of practical applications become possible. These include locomotion or propulsion in soft robotics~\cite{Deng_2020}, precise and repeatable actuation~\cite{Raney2016}, and the reconfigurable devices mentioned earlier~\cite{zareei2020harnessing,jin_2020,Yasuda_2020}. Next steps could include the use of inverse design tools and the controlled use of localized defects to
achieve control of the propagation path or velocity of transition waves, enabling 
more complex actuation and targeted shape changes.

\begin{acknowledgments}
KB, JRR and VT gratefully acknowledge support from NSF grants CMMI-2041410 and CMMI-2041440. Further, KB and  JRR acknowledge support from the Army Research Office grant number W911NF-17–1–0147. Finally, KB acknowledges support from Simons Collaboration on Symmetry-Based Extreme Wave Phenomena.
\end{acknowledgments}

\vspace{0.5cm}

\noindent\textbf{Data availability statement}

Data sharing is not applicable to this article as no new data were created or analyzed in this study.


\bibliography{nonlinear.bib}

\begin{thebibliography}{10}

\bibitem{engheta2006metamaterials}
N.~Engheta and R.~W. Ziolkowski, {\em Metamaterials: physics and engineering
  explorations}.
\newblock John Wiley \& Sons, 2006.

\bibitem{cui2010metamaterials}
T.~J. Cui, D.~R. Smith, and R.~Liu, {\em Metamaterials}.
\newblock Springer, 2010.

\bibitem{craster2012acoustic}
R.~V. Craster and S.~Guenneau, {\em Acoustic metamaterials: Negative
  refraction, imaging, lensing and cloaking}, vol.~166.
\newblock Springer Science \& Business Media, 2012.

\bibitem{deymier2013acoustic}
P.~A. Deymier, {\em Acoustic metamaterials and phononic crystals}, vol.~173.
\newblock Springer Science \& Business Media, 2013.

\bibitem{maldovan2013}
M.~Maldovan, ``Narrow low-frequency spectrum and heat management by
  thermocrystals,'' {\em Physical review letters}, vol.~110, no.~2, p.~025902,
  2013.

\bibitem{BertoldiK.2017Fmm}
K.~Bertoldi, V.~Vitelli, J.~Christensen, and M.~V. Hecke, ``Flexible mechanical
  metamaterials,'' {\em Nature Reviews Materials}, 2017.

\bibitem{KadicMuamer2013Mbe}
M.~Kadic, T.~Bückmann, R.~Schittny, and M.~Wegener, ``Metamaterials beyond
  electromagnetism,'' {\em Rep. Prog. Phys.}, vol.~76, no.~12, 2013.

\bibitem{ChristensenJohan2015Vtfm}
J.~Christensen, M.~Kadic, O.~Kraft, and M.~Wegener, ``Vibrant times for
  mechanical metamaterials (book review),'' {\em MRS Communications}, vol.~5,
  no.~3, pp.~453--462, 2015.

\bibitem{spadaccini2019}
C.~Spadaccini, ``Additive manufacturing and architected materials: New process
  developments and materials,'' {\em Journal of the Acoustical Society of
  America}, vol.~146, no.~4, p.~2756, 2019.

\bibitem{clausen2015}
A.~Clausen, F.~Wang, J.~S. Jensen, O.~Sigmund, and J.~A. Lewis, ``Topology
  optimized architectures with programmable poisson's ratio over large
  deformations,'' {\em Adv. Mater.}, vol.~27, no.~37, pp.~5523--5527, 2015.

\bibitem{raney2015}
J.~R. Raney and J.~A. Lewis, ``Printing mesoscale architectures,'' {\em MRS
  Bull.}, vol.~40, no.~11, pp.~943--950, 2015.

\bibitem{sundaram2019}
S.~Sundaram, M.~Skouras, D.~S. Kim, L.~van~den Heuvel, and W.~Matusik,
  ``Topology optimization and 3d printing of multimaterial magnetic actuators
  and displays,'' {\em Science Advances}, vol.~5, no.~7, p.~eaaw1160, 2019.

\bibitem{LAKES_1987}
R.~Lakes, ``Foam structures with a negative poisson's ratio,'' {\em Science},
  vol.~235, pp.~1038--1040, feb 1987.

\bibitem{Bertoldi2010AdvMat}
K.~Bertoldi, P.~Reis, S.~Willshaw, and T.~Mullin, ``Negative poisson's ratio
  behavior induced by an elastic instability,'' {\em Advanced Materials},
  vol.~22, pp.~361--366, 2010.

\bibitem{Wang_2016}
Q.~Wang, J.~A. Jackson, Q.~Ge, J.~B. Hopkins, C.~M. Spadaccini, and N.~X. Fang,
  ``Lightweight mechanical metamaterials with tunable negative thermal
  expansion,'' {\em Physical Review Letters}, vol.~117, oct 2016.

\bibitem{Boatti_2017}
E.~Boatti, N.~Vasios, and K.~Bertoldi, ``Origami metamaterials for tunable
  thermal expansion,'' {\em Advanced Materials}, vol.~29, p.~1700360, may 2017.

\bibitem{Baughman_1998}
R.~H. Baughman, ``Materials with negative compressibilities in one or more
  dimensions,'' {\em Science}, vol.~279, pp.~1522--1524, mar 1998.

\bibitem{Liu2000}
Z.~Liu, X.~Zhang, Y.~Mao, Y.~Y. Zhu, Z.~Yang, C.~T. Chan, and P.~Sheng,
  ``Locally resonant sonic materials,'' {\em Science}, vol.~289, no.~5485,
  pp.~1734--1736, 2000.

\bibitem{Foehr_2018}
A.~Foehr, O.~R. Bilal, S.~D. Huber, and C.~Daraio, ``Spiral-based phononic
  plates: From wave beaming to topological insulators,'' {\em Physical Review
  Letters}, vol.~120, may 2018.

\bibitem{Ding_2007}
Y.~Ding, Z.~Liu, C.~Qiu, and J.~Shi, ``Metamaterial with simultaneously
  negative bulk modulus and mass density,'' {\em Physical Review Letters},
  vol.~99, aug 2007.

\bibitem{Ma_2018}
J.~Ma, D.~Zhou, K.~Sun, X.~Mao, and S.~Gonella, ``Edge modes and asymmetric
  wave transport in topological lattices: Experimental characterization at
  finite frequencies,'' {\em Physical Review Letters}, vol.~121, aug 2018.

\bibitem{haghpanah2016multistable}
B.~Haghpanah, L.~Salari-Sharif, P.~Pourrajab, J.~Hopkins, and L.~Valdevit,
  ``Multistable shape-reconfigurable architected materials,'' {\em Advanced
  Materials}, vol.~28, no.~36, pp.~7915--7920, 2016.

\bibitem{Rafsanjani_2019}
A.~Rafsanjani, K.~Bertoldi, and A.~R. Studart, ``Programming soft robots with
  flexible mechanical metamaterials,'' {\em Science Robotics}, vol.~4,
  p.~eaav7874, apr 2019.

\bibitem{Raney2016}
J.~R. Raney, N.~Nadkarni, C.~Daraio, D.~M. Kochmann, J.~A. Lewis, and
  K.~Bertoldi, ``Stable propagation of mechanical signals in soft media using
  stored elastic energy,'' {\em Proceedings of the National Academy of
  Sciences}, vol.~113, pp.~9722--9727, aug 2016.

\bibitem{bilal2017}
O.~R. Bilal, A.~Foehr, and C.~Daraio, ``Bistable metamaterial for switching and
  cascading elastic vibrations,'' {\em Proceedings of the National Academy of
  Sciences}, vol.~114, no.~18, pp.~4603--4606, 2017.

\bibitem{jiang2019bifurcation}
Y.~Jiang, L.~M. Korpas, and J.~R. Raney, ``Bifurcation-based embodied logic and
  autonomous actuation,'' {\em Nature communications}, vol.~10, no.~1, p.~128,
  2019.

\bibitem{nesterenko2013dynamics}
V.~Nesterenko, {\em Dynamics of heterogeneous materials}.
\newblock Springer Science \& Business Media, 2013.

\bibitem{Tournat_2010}
V.~Tournat and V.~E. Gusev, ``Acoustics of unconsolidated
  {\textquotedblleft}model{\textquotedblright} granular media: An overview of
  recent results and several open problems,'' {\em Acta Acustica united with
  Acustica}, vol.~96, pp.~208--224, mar 2010.

\bibitem{theocharis2013nonlinear}
G.~Theocharis, N.~Boechler, and C.~Daraio, ``Nonlinear periodic phononic
  structures and granular crystals,'' in {\em Acoustic Metamaterials and
  Phononic Crystals}, pp.~217--251, Springer, 2013.

\bibitem{boechler2011}
N.~Boechler, G.~Theocharis, and C.~Daraio, ``Bifurcation-based acoustic
  switching and rectification,'' {\em Nature Materials}, vol.~10, pp.~665--668,
  2011.

\bibitem{Chong_2017}
C.~Chong, M.~A. Porter, P.~G. Kevrekidis, and C.~Daraio, ``Nonlinear coherent
  structures in granular crystals,'' {\em Journal of Physics: Condensed
  Matter}, vol.~29, p.~413003, sep 2017.

\bibitem{allein2020linear}
F.~Allein, V.~Tournat, V.~Gusev, and G.~Theocharis, ``Linear and nonlinear
  elastic waves in magnetogranular chains,'' {\em Physical Review Applied},
  vol.~13, no.~2, p.~024023, 2020.

\bibitem{Deng_2017}
B.~Deng, J.~Raney, V.~Tournat, and K.~Bertoldi, ``Elastic vector solitons in
  soft architected materials,'' {\em Physical Review Letters}, vol.~118, may
  2017.

\bibitem{Deng_2018NM}
B.~Deng, P.~Wang, Q.~He, V.~Tournat, and K.~Bertoldi, ``Metamaterials with
  amplitude gaps for elastic solitons,'' {\em Nature Communications}, vol.~9,
  aug 2018.

\bibitem{deng2019focusing}
B.~Deng, C.~Mo, V.~Tournat, K.~Bertoldi, and J.~R. Raney, ``Focusing and mode
  separation of elastic vector solitons in a 2d soft mechanical metamaterial,''
  {\em Physical Review Letters}, vol.~123, jul 2019.

\bibitem{deng2019propagation}
B.~Deng, Y.~Zhang, Q.~He, V.~Tournat, P.~Wang, and K.~Bertoldi, ``Propagation
  of elastic solitons in chains of pre-deformed beams,'' {\em New Journal of
  Physics}, vol.~21, p.~073008, jul 2019.

\bibitem{yasuda2019origami}
H.~Yasuda, Y.~Miyazawa, E.~G. Charalampidis, C.~Chong, P.~G. Kevrekidis, and
  J.~Yang, ``Origami-based impact mitigation via rarefaction solitary wave
  creation,'' {\em Science Advances}, vol.~5, p.~eaau2835, may 2019.

\bibitem{Fraternali_2014}
F.~Fraternali, G.~Carpentieri, A.~Amendola, R.~E. Skelton, and V.~F.
  Nesterenko, ``Multiscale tunability of solitary wave dynamics in tensegrity
  metamaterials,'' {\em Applied Physics Letters}, vol.~105, p.~201903, nov
  2014.

\bibitem{Deng_2020}
B.~Deng, L.~Chen, D.~Wei, V.~Tournat, and K.~Bertoldi, ``Pulse-driven robot:
  Motion via solitary waves,'' {\em Science Advances}, vol.~6, p.~eaaz1166, may
  2020.

\bibitem{nadkarni2016unidirectional}
N.~Nadkarni, A.~F. Arrieta, C.~Chong, D.~M. Kochmann, and C.~Daraio,
  ``Unidirectional transition waves in bistable lattices,'' {\em Physical
  review letters}, vol.~116, no.~24, p.~244501, 2016.

\bibitem{vasios2021universally}
N.~Vasios, B.~Deng, B.~Gorissen, and K.~Bertoldi, ``Universally bistable shells
  with nonzero gaussian curvature for two-way transition waves,'' {\em Nature
  communications}, vol.~12, no.~1, pp.~1--9, 2021.

\bibitem{zareei2020harnessing}
A.~Zareei, B.~Deng, and K.~Bertoldi, ``Harnessing transition waves to realize
  deployable structures,'' {\em Proceedings of the National Academy of
  Sciences}, vol.~117, no.~8, pp.~4015--4020, 2020.

\bibitem{jin_2020}
L.~Jin, R.~Khajehtourian, J.~Mueller, A.~Rafsanjani, V.~Tournat, K.~Bertoldi,
  and D.~M. Kochmann, ``Guided transition waves in multistable mechanical
  metamaterials,'' {\em Proceedings of the National Academy of Sciences},
  vol.~117, pp.~2319--2325, jan 2020.

\bibitem{Hwang_2018SR}
M.~Hwang and A.~F. Arrieta, ``Input-independent energy harvesting in bistable
  lattices from transition waves,'' {\em Scientific Reports}, vol.~8, feb 2018.

\bibitem{Yasuda_2020}
H.~Yasuda, L.~M. Korpas, and J.~R. Raney, ``Transition waves and formation of
  domain walls in multistable mechanical metamaterials,'' {\em Physical Review
  Applied}, vol.~13, may 2020.

\bibitem{Grima2000}
J.~Grima and K.~Evans, ``Auxetic behavior from rotating squares,'' {\em J. Mat.
  Sci. Lett.}, vol.~19, p.~1563, 2000.

\bibitem{Cho2014}
Y.~Cho, J.-H. Shin, A.~Costa, T.~Kim, V.~Kunin, J.~Li, S.~Yeon~Lee, S.~Yang,
  H.~Han, I.-S. Choi, and D.~Srolovitz, ``Engineering the shape and structure
  of materials by fractal cut,'' {\em Proc. Natl. Acad. Sci.}, vol.~111,
  p.~17390, 2014.

\bibitem{SShan2015}
S.~Shan, S.~Kang, Z.~Zhao, L.~Fang, and K.~Bertoldi, ``Design of planar
  isotropic negative poisson's ratio structures,'' {\em Extreme Mechanics
  Letters}, vol.~4, p.~96, 2015.

\bibitem{Mullin2007}
T.~Mullin, S.~Deschanel, K.~Bertoldi, and M.~Boyce, ``Pattern transformation
  triggered by deformation,'' {\em Physical Review Letters}, vol.~99,
  p.~084301, 2007.

\bibitem{deng2019anomalous}
B.~Deng, V.~Tournat, P.~Wang, and K.~Bertoldi, ``Anomalous collisions of
  elastic vector solitons in mechanical metamaterials,'' {\em Physical Review
  Letters}, vol.~122, feb 2019.

\bibitem{dauxois2006}
T.~Dauxois and M.~Peyrard, {\em Physics of Solitons}.
\newblock Cambridge University Press, 2006.

\bibitem{Tang_2008}
D.~Y. Tang, H.~Zhang, L.~M. Zhao, and X.~Wu, ``Observation of high-order
  polarization-locked vector solitons in a fiber laser,'' {\em Physical Review
  Letters}, vol.~101, oct 2008.

\bibitem{sen2008solitary}
S.~Sen, J.~Hong, J.~Bang, E.~Avalos, and R.~Doney, ``Solitary waves in the
  granular chain,'' {\em Physics Reports}, vol.~462, no.~2, pp.~21--66, 2008.

\bibitem{Shen2014}
Y.~Shen, P.~G. Kevrekidis, S.~Sen, and A.~Hoffman, ``Characterizing
  traveling-wave collisions in granular chains starting from integrable limits:
  The case of the korteweg--de vries equation and the toda lattice,'' {\em
  Phys. Rev. E}, vol.~90, p.~022905, Aug 2014.

\bibitem{spadoni2010}
A.~Spadoni and C.~Daraio, ``Generation and control of sound bullets with a
  nonlinear acoustic lens,'' {\em Proc. Natl. Acad. Sci.}, vol.~107, no.~16,
  pp.~7230--7234, 2010.

\bibitem{Nesterenko2005}
V.~F. Nesterenko, ``Anomalous wave reflection at the interface of two strongly
  nonlinear granular media,'' {\em Phys. Rev. Lett.}, vol.~95, no.~15, 2005.

\bibitem{Hong2002}
J.~Hong and A.~Xu, ``Nondestructive identification of impurities in granular
  medium,'' {\em Appl. Phys. Lett.}, vol.~81, no.~25, pp.~4868--4870, 2002.

\bibitem{daraio2006tunability}
C.~Daraio, V.~Nesterenko, E.~Herbold, and S.~Jin, ``Tunability of solitary wave
  properties in one-dimensional strongly nonlinear phononic crystals,'' {\em
  Physical Review E}, vol.~73, no.~2, p.~026610, 2006.

\bibitem{nesterenko1983}
V.~F. Nesterenko, ``Propagation of nonlinear compression pulses in granular
  media,'' {\em J Appl Mech Tech Phys}, vol.~24, no.~5, pp.~733--43, 1983.

\bibitem{herbold2013propagation}
E.~B. Herbold and V.~F. Nesterenko, ``Propagation of rarefaction pulses in
  discrete materials with strain-softening behavior,'' {\em Physical review
  letters}, vol.~110, no.~14, p.~144101, 2013.

\bibitem{deng2018effect}
B.~Deng, V.~Tournat, and K.~Bertoldi, ``Effect of predeformation on the
  propagation of vector solitons in flexible mechanical metamaterials,'' {\em
  Physical Review E}, vol.~98, nov 2018.

\bibitem{cherkaev2005transition}
A.~Cherkaev, E.~Cherkaev, and L.~Slepyan, ``Transition waves in bistable
  structures. i. delocalization of damage,'' {\em Journal of the Mechanics and
  Physics of Solids}, vol.~53, no.~2, pp.~383--405, 2005.

\bibitem{truskinovsky2004origin}
L.~Truskinovsky and A.~Vainchtein, ``The origin of nucleation peak in
  transformational plasticity,'' {\em Journal of the Mechanics and Physics of
  Solids}, vol.~52, no.~6, pp.~1421--1446, 2004.

\bibitem{Truskinovsky_2006}
L.~Truskinovsky and A.~Vainchtein, ``Quasicontinuum models of dynamic phase
  transitions,'' {\em Continuum Mechanics and Thermodynamics}, vol.~18,
  pp.~1--21, may 2006.

\bibitem{Truskinovsky_2008}
L.~Truskinovsky and A.~Vainchtein, ``Dynamics of martensitic phase boundaries:
  discreteness, dissipation and inertia,'' {\em Continuum Mechanics and
  Thermodynamics}, vol.~20, pp.~97--122, mar 2008.

\bibitem{Bhattacharya2003}
K.~Bhattacharya, {\em Microstructure of Martensite}.
\newblock Oxford, UK: Oxford University Press, 2003.

\bibitem{porter2009phase}
D.~A. Porter, K.~E. Easterling, and M.~Sherif, {\em Phase Transformations in
  Metals and Alloys, (Revised Reprint)}.
\newblock CRC press, 2009.

\bibitem{falk1984ginzburg}
F.~Falk, ``Ginzburg-landau theory and solitary waves in shape-memory alloys,''
  {\em Zeitschrift f{\"u}r Physik B Condensed Matter}, vol.~54, no.~2,
  pp.~159--167, 1984.

\bibitem{fiebig2016evolution}
M.~Fiebig, T.~Lottermoser, D.~Meier, and M.~Trassin, ``The evolution of
  multiferroics,'' {\em Nature Reviews Materials}, vol.~1, no.~8, p.~16046,
  2016.

\bibitem{zheng2019piezo}
Y.~Zheng, Z.~Wu, X.~Zhang, and K.~Wang, ``A piezo-metastructure with bistable
  circuit shunts for adaptive nonreciprocal wave transmission,'' {\em Smart
  Materials and Structures}, vol.~28, no.~4, p.~045005, 2019.

\bibitem{fang2017asymmetric}
H.~Fang, K.~Wang, and S.~Li, ``Asymmetric energy barrier and mechanical diode
  effect from folding multi-stable stacked-origami,'' {\em Extreme Mechanics
  Letters}, vol.~17, pp.~7--15, 2017.

\bibitem{shan2015multistable}
S.~Shan, S.~H. Kang, J.~R. Raney, P.~Wang, L.~Fang, F.~Candido, J.~A. Lewis,
  and K.~Bertoldi, ``Multistable architected materials for trapping elastic
  strain energy,'' {\em Advanced Materials}, vol.~27, no.~29, pp.~4296--4301,
  2015.

\bibitem{dorin2019vibration}
P.~Dorin, J.~Kim, and K.-W. Wang, ``Vibration energy harvesting system with
  coupled bistable modules,'' in {\em Active and Passive Smart Structures and
  Integrated Systems XII}, vol.~10967, p.~109670G, International Society for
  Optics and Photonics, 2019.

\bibitem{kidambi2017energy}
N.~Kidambi, R.~L. Harne, and K.~Wang, ``Energy capture and storage in
  asymmetrically multistable modular structures inspired by skeletal muscle,''
  {\em Smart Materials and Structures}, vol.~26, no.~8, p.~085011, 2017.

\bibitem{song2019additively}
Y.~Song, R.~M. Panas, S.~Chizari, L.~A. Shaw, J.~A. Jackson, J.~B. Hopkins, and
  A.~J. Pascall, ``Additively manufacturable micro-mechanical logic gates,''
  {\em Nature communications}, vol.~10, no.~1, pp.~1--6, 2019.

\bibitem{Hwang_2018}
M.~Hwang and A.~F. Arrieta, ``Solitary waves in bistable lattices with
  stiffness grading: Augmenting propagation control,'' {\em Physical Review E},
  vol.~98, oct 2018.

\bibitem{hussein2014}
M.~I. Hussein, M.~J. Leamy, and M.~Ruzzene, ``Dynamics of phononic materials
  and structures: Historical origins, recent progress, and future outlook,''
  {\em Applied Mechanics Reviews}, vol.~66, no.~4, p.~040802, 2014.

\bibitem{Nadkarni_2014}
N.~Nadkarni, C.~Daraio, and D.~M. Kochmann, ``Dynamics of periodic mechanical
  structures containing bistable elastic elements: From elastic to solitary
  wave propagation,'' {\em Physical Review E}, vol.~90, aug 2014.

\bibitem{Yasuda_2016}
H.~Yasuda, C.~Chong, E.~G. Charalampidis, P.~G. Kevrekidis, and J.~Yang,
  ``Formation of rarefaction waves in origami-based metamaterials,'' {\em
  Physical Review E}, vol.~93, apr 2016.

\bibitem{frazier2017atomimetic}
M.~J. Frazier and D.~M. Kochmann, ``Atomimetic mechanical structures with
  nonlinear topological domain evolution kinetics,'' {\em Advanced Materials},
  vol.~29, no.~19, p.~1605800, 2017.

\bibitem{Kochmann_2017}
D.~M. Kochmann and K.~Bertoldi, ``Exploiting microstructural instabilities in
  solids and structures: From metamaterials to structural transitions,'' {\em
  Applied Mechanics Reviews}, vol.~69, sep 2017.

\bibitem{Ramakrishnan2020}
V.~Ramakrishnan and M.~J. Frazier, ``Transition waves in multi-stable
  metamaterials with space-time modulated potentials,'' {\em Applied Physics
  Letters}, vol.~117, p.~151901, oct 2020.

\bibitem{nadkarni2016universal}
N.~Nadkarni, C.~Daraio, R.~Abeyaratne, and D.~M. Kochmann, ``Universal energy
  transport law for dissipative and diffusive phase transitions,'' {\em
  Physical Review B}, vol.~93, no.~10, p.~104109, 2016.

\bibitem{mo2019cnoidal}
C.~Mo, J.~Singh, J.~R. Raney, and P.~K. Purohit, ``Cnoidal wave propagation in
  an elastic metamaterial,'' {\em Phys. Rev. E}, vol.~100, p.~013001, 2019.

\bibitem{deng2021dynamics}
B.~Deng, J.~Li, V.~Tournat, P.~K. Purohit, and K.~Bertoldi, ``Dynamics of
  mechanical metamaterials: A framework to connect phonons, nonlinear periodic
  waves and solitons,'' {\em Journal of the Mechanics and Physics of Solids},
  vol.~147, p.~104233, 2021.

\bibitem{Yang_2020}
N.~Yang, C.-W. Chen, J.~Yang, and J.~L. Silverberg, ``Emergent reconfigurable
  mechanical metamaterial tessellations with an exponentially large number of
  discrete configurations,'' {\em Materials {\&} Design}, vol.~196, p.~109143,
  nov 2020.

\bibitem{Yasuda_2020_data}
H.~Yasuda, K.~Yamaguchi, Y.~Miyazawa, R.~Wiebe, J.~R. Raney, and J.~Yang,
  ``Data-driven prediction and analysis of chaotic origami dynamics,'' {\em
  Communications Physics}, vol.~3, sep 2020.

\bibitem{Onorato_2013}
M.~Onorato, S.~Residori, U.~Bortolozzo, A.~Montina, and F.~Arecchi, ``Rogue
  waves and their generating mechanisms in different physical contexts,'' {\em
  Physics Reports}, vol.~528, pp.~47--89, jul 2013.

\bibitem{Chabchoub_2011}
A.~Chabchoub, N.~P. Hoffmann, and N.~Akhmediev, ``Rogue wave observation in a
  water wave tank,'' {\em Physical Review Letters}, vol.~106, may 2011.

\bibitem{Guo2018}
X.~Guo, V.~Gusev, K.~Bertoldi, and V.~Tournat, ``Manipulating acoustic wave
  reflection by a nonlinear elastic metasurface,'' {\em Journal of Applied
  Physics}, vol.~123, p.~124901, 2018.

\bibitem{Guo2019}
X.~Guo, V.~Gusev, V.~Tournat, B.~Deng, and K.~Bertoldi, ``Frequency-doubling
  effect in acoustic reflection by a nonlinear, architected rotating-square
  metasurface,'' {\em Physical Review E}, vol.~99, p.~052209, 2019.

\bibitem{Nassar_2020}
H.~Nassar, B.~Yousefzadeh, R.~Fleury, M.~Ruzzene, A.~Al{\`{u}}, C.~Daraio,
  A.~N. Norris, G.~Huang, and M.~R. Haberman, ``Nonreciprocity in acoustic and
  elastic materials,'' {\em Nature Reviews Materials}, vol.~5, pp.~667--685,
  jul 2020.

\end{thebibliography}
\bibliographystyle{ieeetr}

\end{document}